\documentclass[sigconf,nonacm]{acmart}
\AtBeginDocument{%
  }

\usepackage{balance}
\usepackage{float}
\usepackage{mwe}

\usepackage{siunitx}

\newcommand{\cManual}{MA}
\newcommand{\cManualFull}{\textit{Manual Positioning}\xspace}
\newcommand{\cIcon}{IP}
\newcommand{\cIconFull}{\textit{Situated Icon Preview}\xspace}
\newcommand{\cWindow}{WP}
\newcommand{\cWindowFull}{\textit{Situated Window Preview}\xspace}
\newcommand{\cMiniature}{3D}
\newcommand{\cMiniatureFull}{\textit{3D Preview}\xspace}

\begin{document}

\title[Visualizing Placement Proposals for Window Arrangement in MR]{Visualizing Placement Proposals for Window Arrangement\\in Mixed Reality: A Comparative User Study}

\author{Abdelrahman Zaky}
\orcid{0009-0007-2811-7664}
\affiliation{%
\institution{HCI Group}
  \institution{University of Konstanz}
  \city{Konstanz}
  \country{Germany}}
\email{abdelrahman.zaky@uni-konstanz.de}

\author{Tiare Feuchtner}
\orcid{0000-0002-9922-5538}
\affiliation{%
    \institution{HCI Group}
  \institution{University of Konstanz}
  \city{Konstanz}
  \country{Germany}}
\email{tiare.feuchtner@uni-konstanz.de}

\begin{abstract}
  Adaptive mixed reality (MR) interfaces typically optimize window layouts on behalf of the user, with limited consideration for individual preferences. A promising alternative keeps users in the loop by presenting layout proposals for them to select from, but how these proposals should be visualized remains underexplored. We compare three proposal-visualization techniques for window placement, \cIconFull, \cWindowFull, and \cMiniatureFull, against a \cManualFull baseline. The techniques differ in level of detail and degree of interaction-space context. In a within-subjects user study, 24 participants completed a multi-stage trip-planning task in VR, individually placing seven sequentially introduced windows using each technique. We thereby focus on single-window placement under predefined proposal positions. We measured layouting time, number of layout changes,
  task load, user experience, and preference, complemented by semi-structured interviews. Although \cIconFull and \cWindowFull reduced layouting time compared to \cManualFull — with \cWindowFull also faster than \cMiniatureFull — participants  preferred direct manual control. We discuss the factors shaping this preference (perceived control, cognitive cost, familiarity, and informativeness of the proposal) and outline implications for hybrid approaches, as a promising combination of proposal-based suggestions with manual refinement. 
\end{abstract}

  \begin{teaserfigure}
  \centering
  \includegraphics[scale = 0.567]{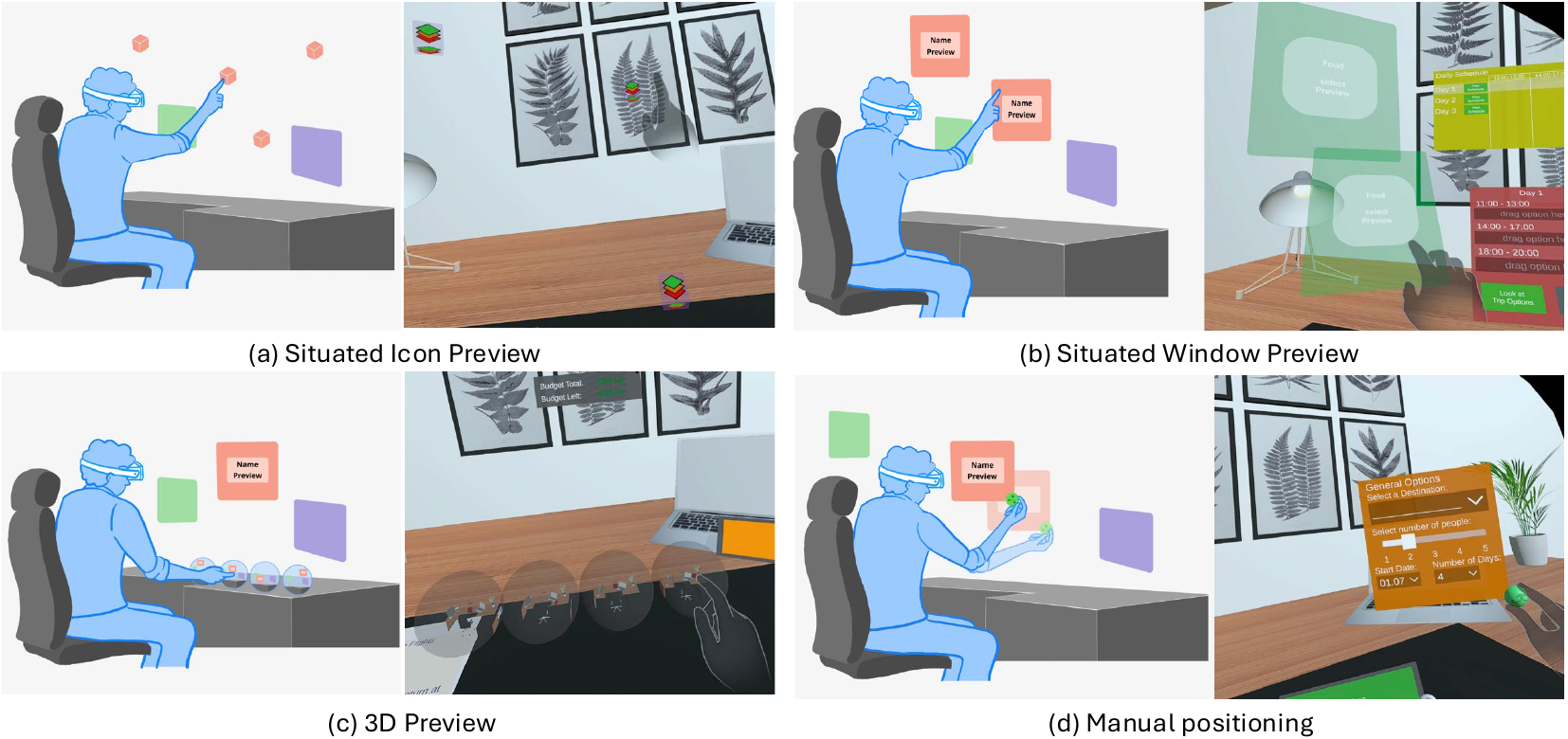}
  \caption{In a comparative evaluation, users completed a task with multiple windows in VR, using four different approaches to adjust their layout: 
    A) \cIconFull, B) \cWindowFull, C) \cMiniatureFull, and D) \cManualFull.}
    \Description{Four views of a person at a desk wearing a VR headset, each demonstrating one window-arrangement technique for placing floating windows: (A) \cIconFull, where small icons mark the candidate positions; (B) \cWindowFull, where preview windows reveal their content at each candidate position; (C) \cMiniatureFull, where a small 3D model of the workspace shows the candidate positions; and (D) \cManualFull, where the user drags a window freely to any position.}
  \label{fig:teaser}
  \end{teaserfigure}

\maketitle

\section{Introduction}
Across a variety of devices and interfaces, users are aided by predefined layouts, grids, tabs, or widgets to organize their content, switch between views, and keep track of open applications.
This is particularly relevant in sensemaking or planning tasks, such as when booking a trip, where many interdependent choices must be made considering many details, from reserving flights and accommodation to scheduling a daily agenda.
Envisioning a future in which our daily work is done in mixed reality (MR) as head-mounted displays (HMDs) extend or replace our current desktop monitors, we must establish window management approaches for 3D interaction spaces that address the inherent challenges: 
In MR applications, users frequently interact with graphical user interfaces (UIs) presented in
of mid-air interaction, also known as 3D UIs. 
Unfortunately, manipulation of such 3D UIs is known to cause muscle strain \cite{bachynskyi2015informing}, the unusual layout of information in 3D space can increase cognitive load \cite{10.1145/3332165.3347945}, and limited tracking volumes hamper the effectiveness of interaction. Further, low tracking accuracy and lack of (haptic) feedback, which may reduce the user's confidence  \cite{10.1145/3334480.3375213}.
Additionally, the placement of 3D UIs may not always align well with the user's activity and environment, potentially leading to safety risks and hindering social interactions \cite{10.1145/3411764.3445349, 10.1145/3526113.3545651}. This underscores a key challenge of MR applications in their ability to adapt to changes in the user's context \cite{10.1145/3332165.3347945}and hinders wider adoption and effective use of the available technologies.

To address such issues, current research has proposed adapting UI elements in MR applications. Different optimization objectives, such as visibility, reachability, and ergonomics \cite{10.1145/3472749.3474750,6948429,10.1145/3411764.3445349,hincapieramos2014consumed}  are commonly considered. To achieve this, multiple algorithms have been developed to solve each objective , for example by formulating placement, level of detail, and application visibility through a mix of rule-based decision-making and combinatorial optimization, as outlined by \citet{10.1145/3332165.3347945}. Further, the context frequently demands multiple simultaneous objectives, requiring multi-objective solvers. 
This has led to the development of toolkits for MR application designers, enabling the combination of multiple objectives during user interface design \cite{10.1145/3526113.3545651}.

In our view, a key remaining challenge of resulting Adaptive MR UIs is that they are adaptive only according to the objectives and strategies originally defined by the application designer. However, should needs change during use, there is commonly no way for the user to modify when and how the adaptations happen. This stands in contrast to systems that offer manual UI adjustment capabilities for the user, such as moving and anchoring of windows to achieve a preferred layout. 
We propose to combine these two approaches, bringing the user into the loop by presenting a set of adaptation possibilities to pick from. Allowing users to select UI layouts that better meet their individual needs can improve the perceived quality and support learning about the trade-offs implicit in the adaptations \cite{10.1145/3544549.3585732}. 
While we expect benefits for system usability when the advantages of fully automated and fully manual adaptation are combined in an adaptation technique \cite{10.1145/3544549.3585732}, it is unclear how the presentation and selection techniques and the choice of algorithm might impact perceived workload.
Arguably, we require further data to make informed decisions how such UI layout proposals should be presented, how many variants to show, how these are selected and how users should be informed about the trade-offs between different objectives.
Our work aims to contribute to the underexplored area of window arrangement for 3D UIs, by comparing different window arrangement techniques and proposal visualization in MR that vary
in visualization size, level of detail for window content, and the degree of interaction space awareness.
Hence, we implemented three visualization techniques: (1) \cWindowFull, (2) \cMiniatureFull, and (3) \cIconFull. 
Focusing on single-window placement,
we evaluate our visualizations in a user study (n=24), comparing these to a baseline with \cManualFull to answer the following research questions:
\begin{enumerate}
\item[\textbf{RQ1}] 
How 
our different designs
of window layout proposals impact interaction performance, user experience and preference?
\item[\textbf{RQ2}] How effective are our window layout proposals compared to the manual placement of windows for UI layout arrangement during a trip planning task?

\end{enumerate}

We contribute (1) an 
empirical comparison of three proposal-visualization techniques for window placement in MR — \cIconFull, \cWindowFull, and \cMiniatureFull — against a \cManualFull baseline.
Our findings offer (2) evidence that users prefer direct manual control over predefined proposals despite the time cost, and reveal four factors that shape this preference: perceived control, cognitive cost of selecting from proposals, familiarity, and proposal informativeness. Based on this, we highlight (3) design implications for hybrid approaches that combine proposal-based suggestions with manual refinement.

\section{Related work}

\subsection{Multiple View Layouts}

Multiple-view (MV) layouts have been extensively studied in 2D
environments \cite{9222323, 9973209, 10.1145/345513.345271}. Shaikh
et al.~\cite{9973209} categorized MV layouts into perception-driven
layouts (flexible, fixed, cascade, focused, split, stacked, and tab)
and content-aware layouts. Roberts et
al.~\cite{https://doi.org/10.1111/cgf.13673} grouped MR view
techniques across 2D interfaces, displays, images, text, and
visualizations. In 3D and MR, layout grids can be
arranged horizontally, vertically, or in circular or cylindrical
forms; Daeijavad and Maurer~\cite{10.1145/3656650.3656756} discussed
the curvature trade-offs for immersive workspaces. Pavanatto et
al.~\cite{9417776} showed that virtual monitors can replace physical
2D screens for productivity work in MR and Spatial
Bar~\cite{10937471} provides thumbnails of open windows to support
switching. Other work integrates MR with interactive
walls~\cite{10.1145/3343055.3359718},
smartphones~\cite{10.1145/3334480.3382812},
smartwatches~\cite{10.1145/2702123.2702331,
10.1145/3544548.3581438}, desktops~\cite{10.1145/3567709, 7469860,
9585781, sidenmark2024desk2desk}, and wall
displays~\cite{10.1145/3359996.3364242, 8781574}. Wen et
al.~\cite{9904883} examined the effect of view layout on situated
analytics in immersive visualization. Lu and
Xu~\cite{10.1145/3491102.3517723} compared three UI transition
mechanisms (low-effort manual, semi-automated, fully-automated),
with the semi-automated condition emerging as both best-performing
and most favored. Despite this breadth, comparative studies of how
layouts should be \emph{visualized} in MR specifically remain
limited.

\subsection{Adaptive MR User Interfaces}

Adaptive MR systems automatically adjust UI placement to the user's context~\cite{10.1145/3332165.3347945, 10322176}. This context spans three dimensions: the \emph{user} (knowledge, abilities, cognitive load, and preferences); the \emph{environment} (the surrounding physical and virtual space, whose geometry may afford or constrain interaction, e.g., haptic feedback from physical surfaces~\cite{10.1145/2047196.2047255}); and the \emph{activity} (the task workflow, intermediate goals, and prerequisites). Common optimization objectives include visibility, reachability, and ergonomics~\cite{10.1145/3472749.3474750, 10.1145/3411764.3445349, 6948429, hincapieramos2014consumed}. Tatzgern et al.~\cite{7504691} adapted AR information density to prevent clutter while supporting iterative exploration. Fender et al.~\cite{10.1145/3173574.3173843} optimized projection-surface visibility by reconstructing the physical space, accounting for user activity. Belo et al.~\cite{10.1145/3411764.3445349} proposed XRgonomics, optimizing 3D-UI ergonomics for mid-air interaction within reach. Lindlbauer et al.~\cite{10.1145/3332165.3347945} introduced context-aware adaptation across multiple objectives simultaneously, and more recent work has extended this to gaze-aware notification placement~\cite{ilo2024goldilocks} and visualization placement for outdoor augmented data tours~\cite{ghaemi2023visualization}. These approaches have been combined into adaptive UI toolkits such as AUIT~\cite{10.1145/3526113.3545651}, which let designers compose multiple objectives during design. Fully-adaptive MR systems, however, give users limited control over when and how adaptations occur. A semi-automated alternative presents a set of Pareto-optimal proposals for the user to choose from~\cite{10.1145/3544549.3585732, 10.1145/3746059.3747645}; how those proposals should be \emph{visualized} for selection has received less attention.

\subsection{Conclusion of Related Work}
Analyzing related work revealed the importance of semi-automated UI layouts over adaptive UI, giving users no options \cite{10.1145/3586183.3606799}. 
It also highlights the importance of investigating how to present and visualize the proposals to the users \cite{10.1145/3544549.3585732}. 
Moreover, it highlights the effect of Multiple View layouts in arranging UI elements, especially windows. It also proposes using different style grids similar to 2D layouts or organizing virtual screens in 3D; however, they deal with grids occluding the environment content without considering the window attributes such as size and functionality.
We address this research gap by comparing different window layout proposal visualizations using various techniques that differ in space, information amount, and point of view. In the next section, we will discuss the visualizations we implemented, explaining their advantages, disadvantages, and design basis.

\section{Supporting Window Arrangement in MR}
\label{section3}
We designed four window layouting techniques informed by prior work, three of which are semi-automated:  with (1) \cIconFull an icon is shown at the center of each proposed window position; with (2) \cWindowFull a full-size preview of the window frame appears at each proposed position; and with (3)  \cMiniatureFull the user sees the proposed window positions within the context of the entire workspace in a world-in-miniature visualization.
Finally, with (4) \cManualFull users can move and place each window freely in the environment by grabbing it with a pinch gesture, which we consider to be a baseline condition. 
 Each of these techniques is described in more detail below.
 
We chose these visualization techniques to cover different aspects of 3D window layout proposal visualizations across three specific dimensions. Firstly, the degree of automation, of which we support two levels: semi-automated or manual placement. We do not include totally-automated placement of windows, as Quentin et al. \cite{10.1145/3290605.3300750} have shown that lack of controllability negatively impacts user satisfaction. Also, their results revealed a strong user preference for manual rectification after automation, even under
high automation accuracy and very poor manual controllability. 
Secondly, the level of detail, or amount of information about the window content presented in the different preview visualizations. This is inspired by the work of Lindlbauer et al.  \cite{10.1145/3332165.3347945} where various levels of detail were explored. Again, we support two levels, high and low. A high level of detail includes the actual window dimensions situated in space and its name, while a low level of detail merely indicates the center position of a window. The third dimension is the degree of interaction space awareness, which can be achieved by changing the viewing perspective from first person to a miniature view. First person perspective allows users to explore proposals that are situated directly in their environment by looking around, while the miniature view offers an overview of the whole 3D world, giving the relative position of the currently previewed window positions within the entire workspace.
To limit the complexity of the layouting problem, we reposition only one window at a time, with our semi-automated techniques offering four position proposals, as described in section \ref{sec:interaction design}. 

\begin{figure*}[h!]
\centering
    \includegraphics[width=\textwidth]{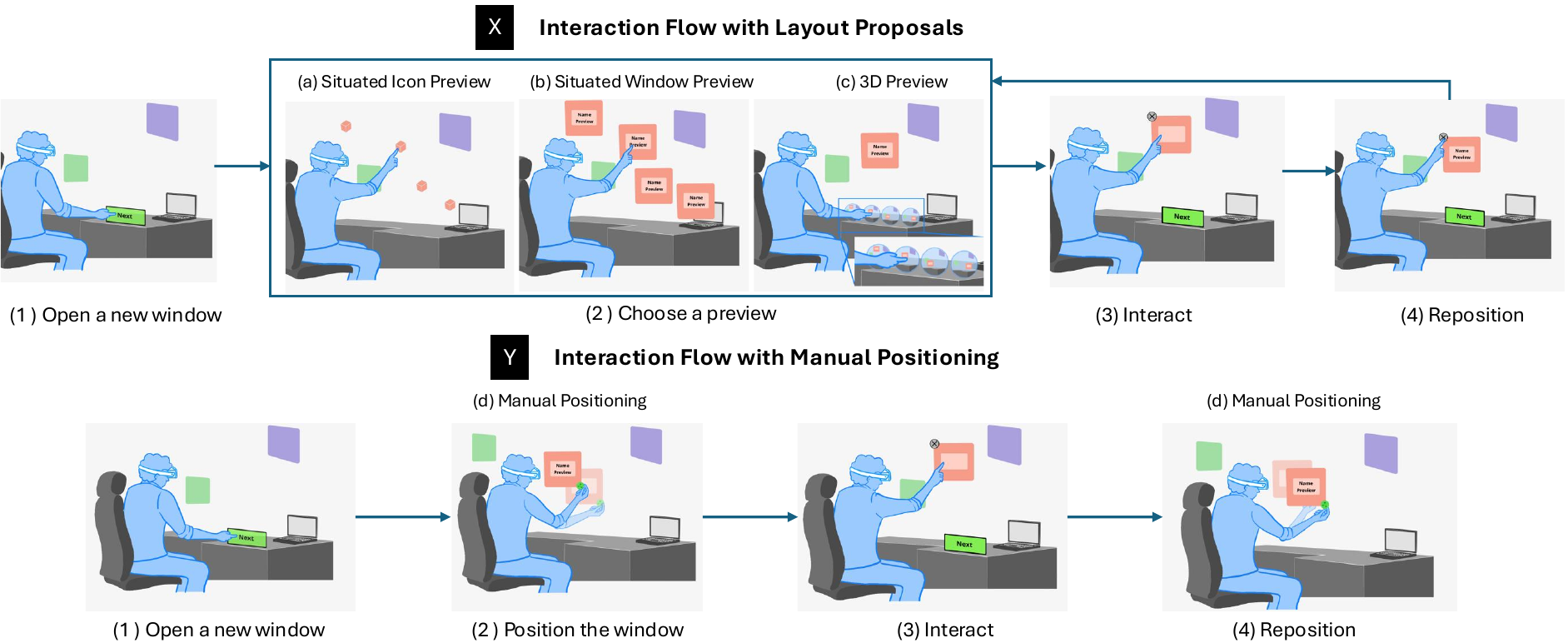}
    \caption{Part X illustrates the interaction flow with the layout proposal techniques a) \cIconFull, b) \cWindowFull, and c) \cMiniatureFull. Part Y shows the interaction flow for \cManualFull. 1. Both parts start with opening a new window. 2. The user then either chooses a preview (X), or manually moves the window (Y). 3. The user interacts with the window content. 4. The user may interrupt their task to readjust the window position, whereby they jump back to (2).}
    \Description{A flow diagram of the interaction with each technique. The upper path (Part X) covers the semi-automated techniques \cIconFull, \cWindowFull, and \cMiniatureFull, and the lower path (Part Y) covers \cManualFull. Both begin with (1) opening a new window; then (2) the user either selects a proposed preview (X) or manually moves the window (Y); (3) works with the window content; and (4) may interrupt the task to readjust the window position, returning to step 2.}
    \label{fig:interactionflow}
\end{figure*}

\subsection{Situated Icon Preview}
Similar to previous work by \citet{10.1145/3544549.3585732}, this technique presents window layout proposals by visualizing an icon at the center point of the proposed UI position  
(\autoref{fig:teaser}, a). 
This is visualized as a small cube with the window icon on each face,
to facilitate good visibility from all perspectives. The icon can be customized, for example to represent the specific window content or application. 
Once a new UI element (window) is opened in the environment, the user can choose one of the multiple proposed positions for it to appear, by tapping on the respective icon. 
Users can also reposition the window at any time by tapping the ``edit button`` attached to the window's corner. This brings back the visualization of position proposals with icons for the user's consideration, who can then can select a new position or revert to the same proposal to end edit mode. After selecting an icon, layouting mode is ended and regular interaction with the windows can resume (\autoref{fig:interactionflow}.X, 4).

This visualization represents the lowest level of detail \cite{10.1145/3472749.3474750} for a window layout proposal in first-person perspective. 
The advantages of this visualization are that (1) the icons are perceived by the user as situated directly in their immediate environment, facilitating body-related judgments (e.g., about visibility and reachability), and (2) the small previews reduce the risk of visual clutter (\autoref{fig:interactionflow}.X, 2-a). Further, this approach benefits from (3) high familiarity, as icons are a typical representation on smartphones and desktop operating systems~\cite{shen2020icon}. On the other hand, users may need to perform a visual search for position proposals, for icons beyond their field of view, and the preview offers no clue about the type or content of the window the user is positioning, or what size it will be. 

\subsection{Situated Window Preview}
Like the previous \cIconFull, this visualization is seen from first-person perspective. Proposals show an empty frame with the size, color, and title of the window to be placed (\autoref{fig:teaser}.b). Our approach is similar to that of \citet{10937471}, which offers a spatial overview of open windows, allowing users to preview their positions and switch between them. 
The richer information in this proposal visualization
may help the user
judge where in the environment this window should be, taking into account potential occlusion of background objects and potentially considering the semantic context of the window in the workflow. 
Further, this technique may benefit from familiar metaphors used in desktop window arrangement systems.
On the other hand,
locating proposals beyond the user's field of view may again require visual search, and the large visualizations can increase visual clutter and occlusion of other content. For example, in \autoref{fig:interactionflow}.X, 2-b, one of the proposals occludes the user's laptop.
Here again, a proposed window position is chosen by selecting the respective preview with a tap. 
As with the \cIconFull, the user can revert to layouting mode at any time during interaction by tapping the ``edit button'' on the window they desire to reposition (\autoref{fig:interactionflow}.X, 4). 

\subsection{3D Previews}

Related work has proposed ``interactive worlds in miniature''~\cite{10.1145/223904.223938} to aid tasks such as object selection and path planning in VR~\cite{BILLINGHURST2001745, danyluk2021design, 9417634}; our technique was primarily inspired by Maslych et al.~\cite{10.1145/3544548.3580873}, who supported the selection of occluded 3D objects via a world-in-miniature preview.  Their mini-map reveals objects that already exist in the scene, whereas we present candidate positions for a window not yet placed.
In our prototype, each of the four layout proposals is presented in a separate sphere containing a scaled model of the workspace (world-in-miniature). 
The window to be positioned is colored red while other objects are gray, and an arrow indicates the user's head-gaze direction. \cMiniature{} spheres can be anchored to any visible element in the environment (in our case, the desk; Figure~\ref{fig:teaser}.c) and are rendered with low opacity to
prevent occlusion of objects behind them. Users select a proposal by tapping its sphere (\autoref{fig:teaser}.c); hovering triggers a temporary situated
preview (\cWindowFull) for first-person verification, and an arrow points to proposals behind the user. The main potential of this
technique lies in contextualizing proposals within the full workspace, which helps organize each window relative to others.
However, we expect that users' unfamiliarity with this technique may make it more difficult to use.

\subsection{Manual Positioning}
\label{sec:manual}
This layouting technique does not fall under the semi-automated window previews and is included as a baseline condition for comparison, as explained in section \ref{sec:study}.
In this technique, 
each opened window is initially centered in the user's field of view  (\autoref{fig:interactionflow}.Y, 2) and can then be
moved and rotated freely, by simply grabbing the manipulation handle -- a sphere attached to the window's bottom right corner -- with a pinch gesture (\autoref{fig:teaser}.d). Visual feedback notifies the user about this implicit switch to layouting mode: the manipulation handle turns blue when touched and green while grabbed. In this mode, the window content becomes non-interactive (i.e., any buttons or sliders cease working) and the position and rotation of the user's hand are applied to the window's transform. Releasing the manipulation handle switches to interaction mode by reactivating the interactive UI elements.
The user can reposition windows at any time during regular interaction (\autoref{fig:interactionflow}.Y, 4).

The main advantage of this technique is the freedom to move a window to any desired position through direct manipulation, which also benefits from great familiarity. However, this may come at the cost of time and muscle strain due to extended or repeated mid-air interaction ~\cite{bachynskyi2015informing,hincapieramos2014consumed}.

\subsection{Considerations for interaction design}
\label{sec:interaction design}
In our sample application, layout adjustments can serve two different goals:
(1) When a new window opens, users must choose its initial position in the environment. 
With the semi-automated techniques, this is done by selecting one of the predefined window positions among the presented proposals. In contrast, in the manual mode the new window appears centered in front of the user and can immediately be dragged away.
(2) Should needs change throughout the workflow,
users can
alter the window layout.
In semi-automated techniques, users must switch to layouting mode by tapping the edit icon (small cross) attached to whichever window they desire to reposition, and then select the preferred proposal. Using the manual technique, users can reposition the window freely by grabbing its manipulation handle. In both cases, this can be repeated until the user finds a suitable window position.

We decided to follow the state of the art established by Johns et al.~ \cite{10.1145/3586183.3606799}, where users position one window at a time, choosing from a set of Pareto-optimal adaptation proposals \cite{10.1145/3544549.3585732}.
Our system can present an arbitrary number of proposals from any source, such as the optimization algorithms of \citet{10.1145/3526113.3545651} or \citet{10.1145/3586183.3606799}, where Pareto-optimal positions are commonly shown as small \cIconFull~\cite{10.1145/3586183.3606799}. To enable a fair comparison against \cWindowFull — where a greater number of larger previews risks clutter — we ran a pilot study with four experienced MR users (three right-handed, one left-handed) to determine (a) the number of proposals and (b) suitable window positions for our trip-planning task.
Working independently in VR, each expert created five non-occluding layouts per window with respect to reachability, dominant-hand placement, and visibility while seated, then convened to discuss these and evaluate different numbers of proposals. All participants reported that more than four proposals caused clutter and overlapped with workspace windows (particularly under \cWindowFull), whereas fewer proposals offered insufficient choice; hence, four proposals balance these constraints. Further, the participants' independent layouts converged on similar positions, supporting their suitability for the task. We extracted the four maximally distant non-occluding positions per window and applied the same predefined set across all three semi-automated conditions, isolating the effect of visualization from that of the placement algorithm. 

\subsection{Design Space}
The three proposal-visualization techniques we examine vary along three design dimensions: (1) the degree of automation, here held constant at semi-automated (proposals plus user selection) but varying across the broader literature from fully manual to fully automated; (2) the level of detail revealed by each proposal, ranging from position-only (\cIconFull) to full content preview (\cWindowFull); and (3) the degree of interaction-space awareness, ranging from local first-person previews (\cIconFull, \cWindowFull) to a world-in-miniature view (\cMiniatureFull). This design space frames our investigation. The empirical study reported next examines how participants respond to combinations of these dimensions, the user-experience factors that emerge, and the implications for hybrid proposal-and-manual systems.

\section{User Study}
\label{sec:study}
    
The study follows a within-subjects design with four conditions: three proposal-visualization techniques (\cIconFull, \cWindowFull, \cMiniatureFull) and a \cManualFull baseline, presented in counterbalanced order. All three semi-automated conditions used the same four predefined proposal positions per window, derived from the pilot study described in \autoref{sec:interaction design}.
Our study design prioritizes controlled comparison over realistic use scenarios in three ways. First, holding the proposal positions constant ensures that observed differences are attributable to visualization rather than to the placement algorithm. Second, participants reposition one window at a time, mirroring single-window adjustment patterns observed in prior work \cite{10.1145/3586183.3606799} and isolating the proposal-visualization comparison from the additional complexity of multi-window coordination. Third, the trip-planning task introduces windows in a fixed sequence, so the comparison reflects the same set of placement decisions across participants and conditions.

\subsection{Participants}

We recruited 24 participants (12 male, 11 female, 1 non-binary), aged 20–38 years (M = 27.29, SD = 4.83), through convenience sampling via institutional mailing lists, social media groups, and campus posters. Inclusion criteria were a minimum age of 18 and no severe history of VR sickness; no exclusion criteria were applied. The sample comprised 18 undergraduate students, 3 PhD students, 2 researchers, and 1 software developer. All participants had normal or corrected-to-normal eyesight. On a scale of 1 (``excellent'') to 5 (``weak''), they reported very good English reading and understanding (M = 1.83, SD = 1.37). Twenty-two were right-handed, and two were left-handed. On a scale of 1 (``never'') to 7 (``extensive''), participants reported moderate-to-high VR/AR experience (M = 3.71, SD = 2.15), low-to-moderate 3D user interface interaction (M = 3.16, SD = 2.15), and high usage of 2D window management tools (M = 4.87, SD = 1.87).
 The user study
was conducted under blanket approval by the ethics committee at the University of Konstanz.
All participants provided written informed consent before participation, and were compensated at the local minimum wage.

\subsection{Apparatus}

Participants used a Meta Quest Pro (2023, 722 grams),
which was connected to a PC with a link cable with the following specifications: CPU: Intel i9 9900K, RAM: 32 GB, GPU: Nvidia RTX 3090. It was also used to complete various questionnaires. 
The VR application was implemented in Unity 2022.3.8f1 \cite{unity}.
Hand tracking is performed via the Quest Pro's onboard cameras. We detect finger pokes for button-like UI elements and pinch gestures for direct manipulation (grabbing, moving, releasing) of windows and proposals.

\subsection{Procedure}
\begin{figure}[h!]
\centering
    \includegraphics[width= 0.4\textwidth]{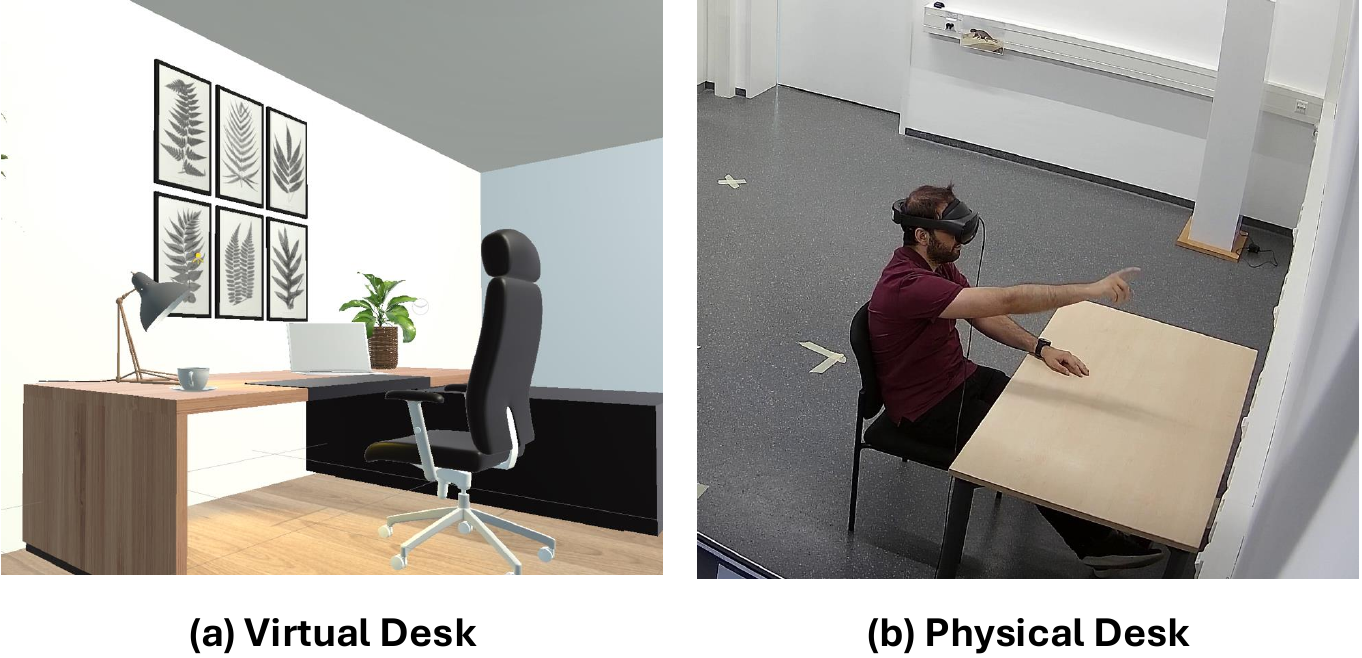}
    \caption {Participants were seated at a desk in an office space (left), while in VR, a virtual office was presented (right).}
    \Description{Two images side by side: on the left, a participant sits at a desk in a physical office wearing a VR headset; on the right, the matching virtual office shown inside the headset.}
    \label{fig:desk}
\end{figure}

Each session lasted approximately 90 minutes. Participants were first welcomed, introduced to the study's purpose, given the opportunity to ask questions, and asked to sign a consent form ($\approx$ 10 min). They then completed a demographic questionnaire on a PC ($\approx$ 5 min). Throughout the session, participants were seated at a desk in a physical workspace mirrored in VR (\autoref{fig:desk}). The four conditions — \cManualFull, \cIconFull, \cWindowFull, and \cMiniatureFull — were presented in an order determined by a full Latin square across the 24 participants, ensuring each condition appeared equally often in each ordinal position.

For each condition, participants first completed a brief training session ($\approx$3 min): planning a one-day trip on the virtual desk using the assigned technique, with the experimenter present to answer questions. They were asked to prioritize the window arrangement over the trip-planning content while still following the trip details to progress through the task, and to create a clear layout in which all windows were visible and accessible. Once comfortable with the technique, participants began the recorded task: planning a three-day trip to a different city using the same technique ($\approx$ 10 min), after which they completed the UEQ and NASA TLX ($\approx$ 3 min). This sequence was repeated for the remaining three conditions. Each condition used a different city, as did the training task, to avoid learning effects from task repetition; tasks across conditions were matched in difficulty, expected number of interactions, and trip length (three days). At the end of the session, participants rated the four techniques on personal preference and took part in a semi-structured interview about their experience ($\approx$ 15 min).

\subsection{Task}
\label{sec:task}
\begin{figure*}[h!]
\centering
    \includegraphics[width= \textwidth]{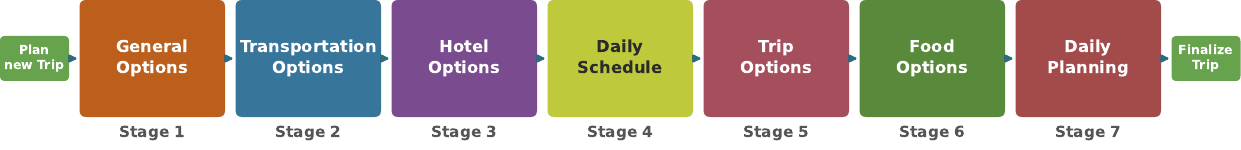}
   \caption{Abstract overview of the seven trip-planning task stages. The actual windows in full detail are shown in \autoref{app:taskflow}.}
    \Description{A horizontal sequence of the seven trip-planning task stages, each a coloured window connected by arrows: General Options, Transportation, Hotel, Daily Schedule, Trip Options, Food Options, and Daily Planning. A green ``Plan new Trip'' button starts the sequence and a green ``Finalize Trip'' button ends it, with each stage numbered Stage 1 to Stage 7.}
    \label{fig:taskflow}
\end{figure*}

We designed a trip-planning task involving seven windows, introduced
across seven sequential stages, illustrated in \autoref{fig:taskflow}. 
 \autoref{app:taskflow} shows the windows in full detail for each stage. The task covers planning transportation, accommodation, daily schedule, and food options, and is structured to elicit multiple window-layout decisions per session. Window interactions include drop-down menus, scrolling, check boxes, sliders, search boxes, and buttons. At each stage, prior windows remain in the environment, so participants progressively construct a multi-window layout.

The task idea is guided by Lindlbauer et al. ~\cite{10.1145/3332165.3347945}. Each stage of the task is inspired by existing applications: browsing for and comparing flights (Google Flights~\cite{googleflight}), searching for all types of transportation (Omio~\cite{omio}), finding suitable accommodation (Booking~\cite{booking}, Airbnb~\cite{airbnb}),
and planning vacation days with different activities, events, and meals (Wanderlog~\cite{wanderlog}). Grounding each stage in an existing application keeps the task realistic while requiring no specialized knowledge from participants.
  
We chose this task because it inherently involves many concurrent windows: a desktop or mobile equivalent would require juggling browser tabs or apps to compare flights, accommodations, daily schedules, and budgets, resulting in going back and forth between them. Therefore, in our task, participants encounter an ``Out of Budget'' error and must adjust some aspects they have already chosen, such as the number of days, the number of people, or the flights. We verified that every participant encountered this error in all tasks, and that the provided instructions reliably resolved it.

The window arrangement task is simultaneous to the main task and includes two main parts: (1) Initial layout: this happens when there is a new window after participants finish a specific stage. (2) Modifying the position of a window by selecting the modify icon on any window to reposition it to change the window layout. In general, participants get to arrange one window at a time. However, they can perform the rearrangement any number of times until they have the window layout they want. Both steps differ in the visualization and the manipulation according to the layouting technique, which is explained in  \ref{section3}.

\subsection{Dependent Variables}

We measured layout-task completion time, the number of layout changes per task, and overall task-completion time. Subjective workload was assessed with the unweighted NASA TLX~\cite{HART1988139}, user experience with the UEQ~\cite{laugwitz2008construction}.
Both questionnaires are validated and widely used in MR research.
Further, preference was rated on a 1--5 scale (one = worst, five = best; ties permitted).
A semi-structured interview elicited the reasons for each rating, perceived advantages and disadvantages of each technique, and participants' ideas about proposal positions, visualizations, and combinations of techniques.

\section{Main Study Findings}
\label{sec:findings}

 In this section, we present the main findings of our lab study, focusing on significant effects; the full pairwise comparison tables underlying every reported omnibus are provided in
Appendix~\ref{app:pairwise}. The quantitative data were analyzed using IBM SPSS 29~\cite{spss}. For continuous data, normality was assessed using the Shapiro-Wilk test; if normality held, we used one-way repeated-measures ANOVA; otherwise, Friedman's ANOVA. Ordinal data (questionnaire ratings and preferences) were analyzed 
non-parametrically 
(Friedman's ANOVA) by default. Post-hoc pairwise comparisons used Dunn's test (for Friedman) or pairwise estimated marginal means contrasts (for ANOVA), in both cases with Bonferroni adjustment. The $p$-values reported throughout this section are the Bonferroni-adjusted values and are tested against $\alpha = .05$. A post-hoc sensitivity analysis indicating the minimum effect sizes detectable with our design is reported in \textit{Appendix~\ref{app:poweranalysis}}.

Qualitative findings were derived from the semi-structured 
interviews using reflexive thematic analysis~\cite{braun2006using}, 
with a primary coder and a validating second coder; the detailed procedure is reported in \autoref{app:thhematicprocedure}. In the following, we succinctly indicate conditions using the subscripts 
$_{\cManual}$ for \cManualFull, 
$_{\cIcon}$ for \cIconFull,
$_{\cWindow}$ for \cWindowFull,
and $_{\cMiniature}$ for \cMiniatureFull.

A trial-order analysis verifying that learning effects do not confound the condition-level comparisons is reported in \autoref{app:learningeffect}.

\subsection{Performance}
\begin{figure*}
\centering
    \includegraphics[width=0.905\textwidth, angle=0]{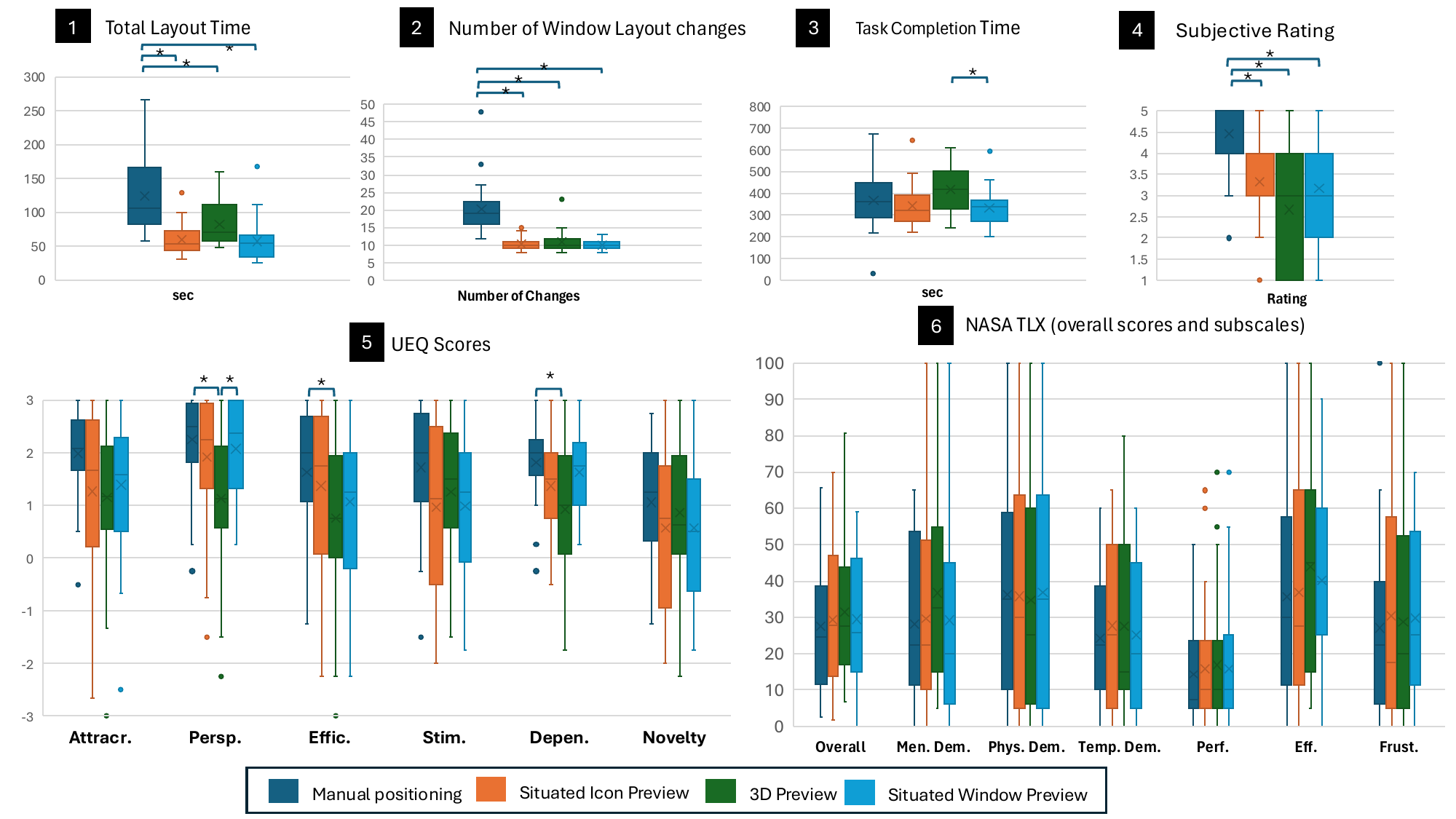}
    \caption{(1-3) Boxplots for performance measure (1) total layouting time, (2) number of window layout changes, and (3) task completion time. (4) subjective rating, (5) UEQ scores, and (6) workload measures: overall scores of NASA TLX and its subscales.}    
    \Description{A grid of boxplots comparing the four techniques (\cIconFull, \cWindowFull, \cMiniatureFull, and \cManualFull) across six measures: (1) total layouting time, (2) number of window layout changes, (3) task completion time, (4) subjective preference rating, (5) UEQ scores, and (6) NASA-TLX overall workload with its subscales. Each panel shows one box per technique with medians, quartiles, and outliers.}
    \label{fig:boxplots}
\end{figure*}

\subsubsection{Layouting time}
\label{sec:layoutingTime}
Layouting time per condition, shown in \autoref{fig:boxplots}.1, was
compared using a Friedman's ANOVA, which revealed significant differences
between conditions (x$^2$(3) = 32.45,  $p< .001$, W = .45). Post-hoc pairwise comparisons showed that participants took significantly more time to layout the windows with \cManualFull
($M_{\cManual}=\qty{120.33}{\second}$, $SD = \qty{52.70}{\second}$)
than with \cIconFull
($M_{\cIcon}=\qty{58.83}{\second}$, $SD = \qty{24.09}{\second}$,
$z = 4.36$, $p < .001 $, $r = .63$), and \cWindowFull
($M_{\cWindow}=\qty{54.54}{\second}$, $SD = \qty{31.82}{\second}$,
$z = 5.26$, $p < .001$,  $r = .76$). Layouting was also significantly
faster with \cWindowFull than with \cMiniatureFull
($M_{\cMiniature}=\qty{79.40}{\second}$, $SD = \qty{28.98}{\second}$,
$z = 2.79$, $p = .031$, $r = .40$). No significant differences were found among the remaining contrasts.

\subsubsection{Overall Task Completion Times}
\label{sec:overallanalysis}
The overall completion time for the trip planning task per condition,
shown in \autoref{fig:boxplots}.3, is the duration from when the
participant selected ``Plan New Trip'' until ``Finalize Trip''
(see \autoref{fig:taskflow}). A repeated-measures ANOVA revealed a
significant difference between the four conditions ($F(3,69) = 3.94$,
$p = .012$, $\eta_p^2 = .15$), with pairwise comparisons indicating that
participants were faster with \cWindowFull
($M_{\cWindow} = \qty{321.91}{\second}$, $SD = \qty{69.18}{\second}$)
than with \cMiniatureFull ($M_{\cMiniature} = \qty{412.18}{\second}$,
$SD = \qty{108.70}{\second}$, $\mathit{MD} = \qty{-90.27}{\second}$,
$p = .013$). No further significant differences were found between the
remaining conditions.

\subsubsection{Number of Layout changes }
\label{sec:layoutChanges}
The total number of layout changes per task, illustrated in
\autoref{fig:boxplots}.2, is counted as the number of times participants switched into layouting mode: for proposal conditions, this happened at initial appearance of each window and then upon tapping the
``edit button'' on a window, while for the manual condition, when grabbing a window's manipulation handle. 
A Friedman's ANOVA revealed significant differences between conditions
(x$^2$(3) = 43.05,  $p < .001$, $W = .60$), 
with more window layout changes done in 
\cManualFull ($M_{\cManual}={19.17}$, $SD = {4.91}$)
in pairwise comparison to 
\cIconFull ($M_{\cIcon}={10.17}$, $SD = {2.01}$,
$z = 5.03$, $p < .001 $, $r = .73$), 
\cWindowFull 
($M_{\cWindow}={9.96}$, $SD = {1.55}$,
$z = 5.37$, $p < .001$, $r = .77$),
and \cMiniatureFull 
($M_{\cMiniature}={10.50}$, $SD = {2.32}$,
$z = 5.03$, $p < .001$, $r = .73$). 
Layout changes in the three proposal techniques were similar.

\subsubsection{Number of Layout Changes across Stages}

\begin{figure}[t]
 \centering
     \includegraphics[scale=0.3]{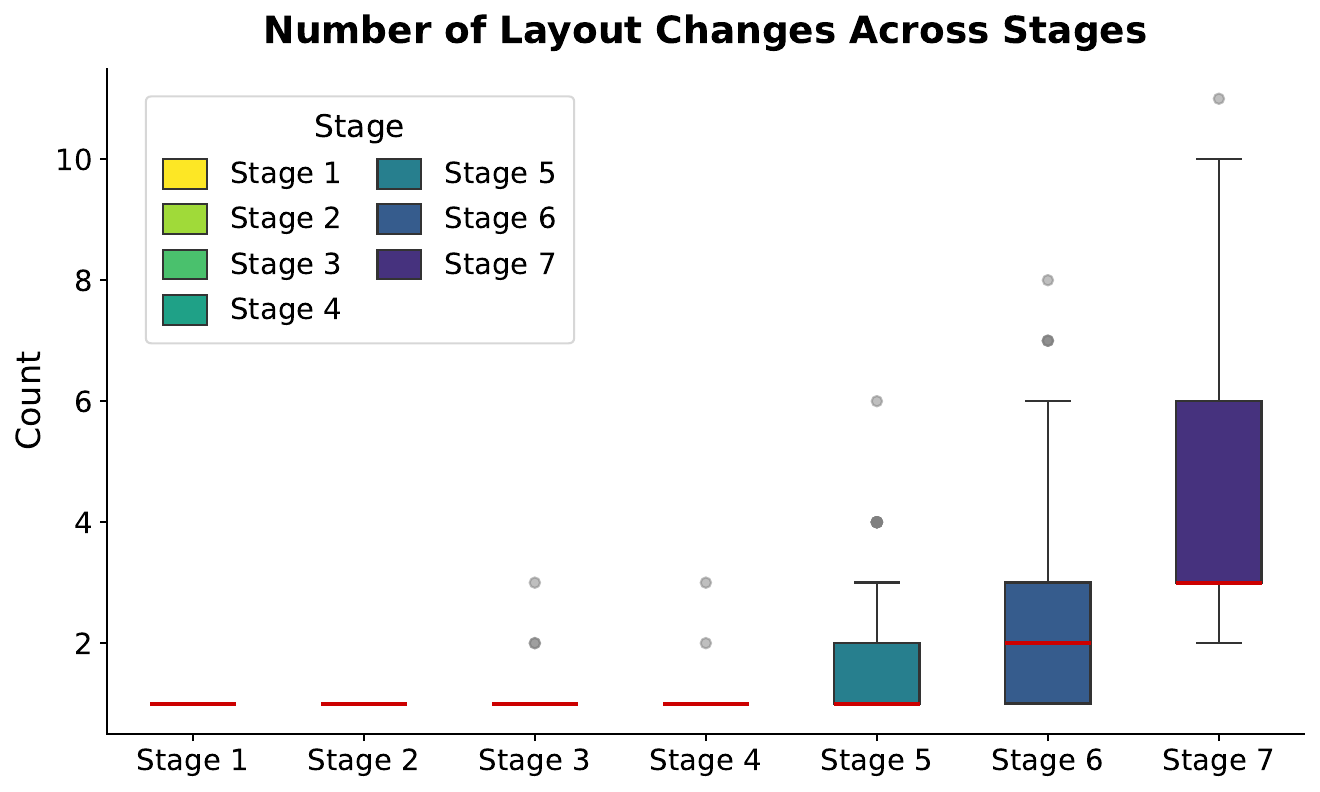}
     \caption{Boxplot illustrating the number of layout changes across different stages of the task}
     \Description{A boxplot of the number of window layout changes at each successive stage of the trip-planning task, one box per stage, showing that readjustments were concentrated in the later stages.}
     \label{fig:stages}
 \end{figure}

 The task is divided into 7 stages, with a new window introduced in each stage. To test whether layout activity varied across stages, independent of condition, we pooled the four sessions per participant, treating each session as the unit of analysis (N = 96 sessions: 24 participants × 4 conditions), as illustrated in \autoref{fig:stages}. A Friedman's ANOVA indicated a significant effect of stage on the number of layout changes  (x$^2$(6) = 477.07,  $p < .001$, $W = .83$). Pairwise comparisons showed that stages 6 ($Mdn = 2.44$) and 7 ($Mdn = 4.48$) had significantly more layout changes than each of stages 1 through 5 (all $|z|$ $\geq$ $4.51$, all $p < .001$ after Bonferroni correction); stage 7 also had significantly more changes than stage 6 ($z =$ -5.90, $p < .001$). The earlier stages 1–5 did not differ significantly from one another. 

\subsection{User Experience}
We assessed subjective user experience using three instruments: NASA TLX  \cite{HART1988139} for perceived workload, the User Experience Questionnaire (UEQ) \cite{laugwitz2008construction} for perceived hedonic and pragmatic quality, and a 5-point preference rating.

\subsubsection{Workload}
According to scores on the NASA TLX questionnaire, participants’ task load was low overall (see  \autoref{fig:boxplots}.6). Friedman's ANOVA shows no significant differences between conditions in the overall score (x$^2$(3) =  3.41,  $p = .332$, $W = .05$). 

\subsubsection{User Experience Questionnaire (UEQ)}
For UEQ scores, visualized in \autoref{fig:boxplots}.5, a Friedman's ANOVA revealed significant differences in terms of Dependability  (x$^2$(3) =  10.82, $p = .013$, $W = .15$), whereby in posthoc comparisons \cManualFull ($M_{\cManual}={1.81}$, $SD={.73}$) was rated higher than \cMiniatureFull ($M_{\cMiniature}={.93}$, $SD={1.24}$, $z = 3.02$, $p = .015$, $r = .44$). 
Efficiency also differed significantly (x$^2$(3) =  11.28,  $p = .010$, $W = .16$), with \cManualFull ($M_{\cManual}={1.62}$, $SD={1.24}$) again rated higher than \cMiniatureFull ($M_{\cMiniature}={.75}$, $SD={1.53}$, $z = 2.85$, $p = .026$, $ r = .41$). 
Regarding Perspicuity, there was a significant difference (x$^2$(3) =  17.43,  $p < .001$, $W = .24$), with \cMiniatureFull ($M_{\cMiniature}={1.12}$, $SD={1.25}$) being rated significantly worse than both \cWindowFull 
($M_{\cWindow}={2.07}$, $SD={.88}$, $z = -2.96$, $p =.018$, $r = .43$) and \cManualFull($M_{\cManual}={2.24}$, $SD={.83}$, $z = 3.75$, $p = .001$, $r = .54$). 
Omnibus differences in terms of Stimulation (x$^2$(3) =  8.12,  $p = .044$) and Attractiveness (x$^2$(3) =  9.41,  $p = .024$, $W = .13$) did not survive pairwise Bonferroni correction.
Finally, there was no significant effect of conditions on Novelty (x$^2$(3) =  4.01,  $p = .260$, $W = .06$); pairwise tables for the significant subscales are reported in Appendix~\ref{app:b-ueq}.

\subsubsection{Subjective Rating}
\label{sec:subjectiveRating}
We asked participants to rate each condition based on their preference on a scale from one (worst) to five (best). As visualized in \autoref{fig:boxplots}.4, a Friedman's ANOVA revealed a significant difference between conditions (x$^2$(3) = 20.20, $p < .001$, W = .28), whereby \cManualFull ($M_{\cManual} = 4.46$, $SD = 0.93$) was preferred over all remaining conditions: \cMiniatureFull ($M_{\cMiniature} = 2.67$, $SD = 1.37$, $z = 3.91$, $p = .001$, $r = .57$), \cIconFull ($M_{\cIcon} = 3.33$, $SD = 1.17$, $z = 3.24$, $p = .007$, $r = .47$), and \cWindowFull ($M_{\cWindow} = 3.17$, $SD = 1.20$, $z = 3.13$, $p = .010$, $r = .45$). There were no further significant differences in preference between proposal conditions. 

In additional exploratory analysis, we found that preference correlated with prior experience (\autoref{app:familiarity}): preference for \cManualFull{} over the proposal techniques tended to increase with 2D window-management experience, whereas preference for \cIconFull{} decreased with greater VR/AR and 3D-UI experience.
\subsection{Semi-Structured Interview}  
\label{sec:interview}

Four themes emerged from the analysis: \textit{Perceived Control 
over Placement}, \textit{Cognitive Cost of Selecting from Proposals},
\textit{Familiarity and Visual Recognition}, and \textit{Informativeness
of the Proposal} (see \autoref{app:thhematicprocedure} for the detailed procedure of the thematic-analysis).

\textbf{\textit{Theme \#1: Perceived Control over Placement}}.
Participants placed strong value on manual control over where windows ended up and, conversely, experienced the proposal techniques as constraining, even when they appreciated their efficiency. A common characterization of \cManualFull was that ``Manual Placement has more freedom'' (P2): 19 participants expressed wanting to ``put it wherever I want'' (P4), and 15 valued fine-grained adjustments. The constraint of proposal techniques was felt acutely: e.g., P1 reported, ``In proposal conditions, I sometimes wished to place a window at a certain position'', a sentiment shared by 11 participants under the ``restrictive'' code. The most-endorsed code in the dataset (n = 20) was a desire to combine techniques, as P2 described, ``you can choose the predefined, you can adjust it if you want, but it doesn't restrict you''. This suggests that for proposal techniques, it is critical to support manual control for fine-tuning the window placement. 

\textbf{\textit{Theme \#2: Cognitive Cost of Selecting from Proposals}}.
Selection effort varied substantially across the three proposal techniques and influenced participants' experiences. \cIconFull was described as the lowest-effort option by 13 participants, with P3 capturing the typical reaction: ``It's easy, it's intuitive. It's pretty simple''. \cMiniatureFull, in contrast, required the highest effort; 8 participants explicitly cited its multi-step interaction, which P6 articulated as: ``It takes some time to preview and see where it is, and then decide. It's like two steps, then just one''. This pattern explains why \cMiniatureFull led to the highest interaction times and was lowest-rated in preference, despite participants finding it visually engaging.

\textbf{\textit{Theme \#3: Familiarity and Visual Recognition}}.
The recognition of the visual representation shaped how readily participants engaged with each proposal type. The cube icon in \cIconFull was criticized as semantically uninformative — P2 stated plainly, ``I didn't understand what the icon was supposed to represent''. 
In contrast, several participants found  \cMiniatureFull visually engaging — P3: ``I think it's very creative. I like the visualization in very small detail with the desk and everything.'' 
Unfortunately, engagement did not translate into preference, as the cognitive cost of using this technique (Theme 2) outweighed its novelty appeal. This suggests that
conceptual familiarity is key for successful adoption.

\textbf{\textit{Theme \#4: Informativeness of the Proposal}}.
 Participants articulated a clear trade-off between proposal richness and visual clutter. \cWindowFull was valued for showing the actual window dimensions and content at the proposed location; 9 participants appreciated that it was ``exactly how it will look like'' (P19), and P9 explained the comparative advantage: ``\cWindowFull is a little bit better than Icon Previews because you know which size the items will be''. Conversely, the minimal representation of \cIconFull left participants unable to anticipate the outcome: ``I don't know what the window will look like. Don't know the size of the window'' (P18). However, the more informative representations of \cWindowFull also produced visual clutter, particularly when multiple proposals were visible at once. As P12 put it, ``I didn't like the \cWindowFull, mostly because of the visual clutter''. Together, these patterns indicate that informativeness is a context-dependent dimension,
 and the visualization itself competes for attention with the workspace.

 \textit{\textbf{Beyond the four themes: a personalization signal.}}
A separate pattern emerged outside the four themes: 18 participants endorsed configuring reusable layouts for repetitive tasks, framing personalization as a complement to proposal selection rather than a replacement. 
P11 captured the typical framing: ``I would appreciate configuring my layout if I use it every day.'' This pattern suggests that even users who appreciate proposal-based selection still want a separate mechanism to codify recurring workflows.

\section{Discussion}
\label{sec:discussion}

Our findings highlight a preference–efficiency trade-off: \cManualFull required significantly more layouting time and roughly twice as many layout changes as \cIconFull or \cWindowFull, yet was strongly preferred over all three proposal techniques.
However,  the increased layouting effort does not appear to have impacted overall task completion time with \cManualFull — likely because layouting activity was mostly limited to the final two task stages. Hence, the penalty for practical efficiency may not be severe enough to significantly impact users' experience. 
Among the proposal techniques, \cWindowFull outperformed \cMiniatureFull in both layouting time and overall task completion, and here \cMiniatureFull was also rated lower on multiple UEQ subscales.
\cIconFull and \cMiniatureFull, in contrast, produced statistically identical distributions of layout changes ($z = 0.000$): participants required the same number of re-layout adjustments under both conditions. \cMiniature's longer layouting time therefore reflects the cost of its initial selection.

Our finding that \cManualFull was preferred over proposal techniques, despite requiring more effort, is consistent with \citet{10.1145/3290605.3300750} reporting that controllability outweighs automation accuracy in user preference,
whereby we extend these findings from 2D desktop tasks to multi-window placement in immersive MR.
 Similarly, recent work on VR authoring reports manual specification affording the greatest sense of agency~\cite{zhang2024vrcopilot}.
 The broader pattern that users favor direct manipulation even when adaptable alternatives are more efficient has been documented across adaptive vs. adaptable menus \cite{10.1145/985692.985704}; our results suggest that this also holds when the ``adaptive'' alternative is a semi-automated proposal-selection mechanism that keeps the user in the loop.

Prior work on semi-automated MR layout has primarily explored isolated proposal techniques: Johns et al. \cite{10.1145/3544549.3585732, 10.1145/3586183.3606799} introduced icon-based proposals over Pareto-optimal positions, Pavanatto et al.'s Spatial Bar \cite{10937471} provides thumbnail-based window previews, and \citet{10108428} explored world-in-miniature representations for occluded object selection. Inspired by these alternative representation methods, we compare three visualizations and show that visualization choice substantively shapes outcomes.
We further complement recent work on preference-guided proposal selection \cite{10.1145/3746059.3747645}, which optimized the selection of proposed positions, by highlighting that the \textit{manner} in which positions are presented is an equally consequential design choice. Concurrent work by \citet{luospatial} on AR document arrangement under predefined layouts and by \citet{pavanattowindow} on window-management strategies for virtual displays corroborates the broader pattern: visualization design shapes user experience in spatial information work.

Fully-automated approaches for UI layout in MR, such as Lindlbauer et al.'s context-aware adaptive UIs \cite{10.1145/3332165.3347945} and Belo et al.'s ergonomic optimization toolkits \cite{10.1145/3411764.3445349, 10.1145/3526113.3545651}, aim to place windows
according to objectives like reachability, visibility, and ergonomics. Recent work sharpens this paradigm in two directions: Li et al.'s SituationAdapt \cite{lisitiuationadapt} uses LLM reasoning to incorporate situational context (e.g., the presence of others, shared displays); and concurrent theoretical work by \citet{10.1145/3544549.3585732} shows that fully automated optimization is limited. 
The findings presented in our paper aim to provide complementary empirical grounding for the latter concern: even when users were given a selection step with proposals and the system did not fully automate the decision, our study participants strongly preferred manual control. This suggests that improvements in fully automated accuracy alone, including LLM-based contextual reasoning, may not be sufficient to overcome the controllability preference documented in \citet{10.1145/3290605.3300750}, which we confirm. Future fully automated MR systems should therefore not only aim to accurately predict the user's intended layout but also to preserve the user's sense of agency when presenting such predictions.

Our qualitative analysis identified four factors that shape technique preference, plus further concerns about position suitability and user-defined personalization (n=18). Perceived control explains the preference pattern: 19 participants
valued the freedom of placement, and 15 emphasized the importance of fine adjustment, while the most-endorsed code (n=20) was the desire to combine proposal selection with manual refinement, mirroring the controllability-over-accuracy preference \cite{10.1145/3290605.3300750}. Cognitive cost may explain why \cMiniatureFull consistently underperformed, with 8 participants remarking on its multi-step interaction, while physical effort played a negligible role as this was comparable in \cMiniatureFull and \cManualFull. Further, familiarity and visual recognition are critical: \cMiniatureFull was novel and visually engaging but did not translate into preference, whereas the preference for \cManualFull may partly reflect legacy bias, as the technique resembles familiar 2D interaction. Finally, informativeness is governed by a trade-off: \cWindowFull's content-revealing previews informed placement decisions but produced visual clutter at scale.Conversely, \cIconFull{}'s minimal representation was rated lowest by the most VR/AR- and 3D-UI-experienced participants (\autoref{app:familiarity}), suggesting they found it least informative. Beyond visualization, participants themselves assessed the suitability of the proposed positions: 20 participants confirmed positions were reachable and visible, and several suggested algorithmic improvements, including placement in relation to the dominant-hand, semantic window grouping \cite{10.1145/3472749.3474750, 10.1145/3544548.3580873}, grid arrangements around the user, and projection onto suitable physical objects \cite{10.1145/3626472}.

Several design implications follow. Most concretely, hybrid systems combining proposal-based selection with manual refinement show promise. Layout management appears most valuable for tasks with more than five windows; in our study, readjustments most often occurred in the final two stages. For visualization design, (1) minimize selection steps and ensure adequate proposal size for visibility when designing a \cMiniatureFull (e.g., a world-in-miniature); and (2) cap the number of simultaneously visible proposals for \cWindowFull to alleviate clutter.

\section{Limitations and Future Work}
\label{sec:limitations}

Our study deliberately constrains the rearrangement problem along three dimensions, with implications for how our findings transfer. First, participants arrange one window at a time, while real-world MR usage often involves rearranging multiple windows together. 
However, simultaneous multi-window rearrangement poses additional challenges for spatial memory \cite{10.1145/3544548.3581438} and computational complexity, and may interact with proposal visualization in ways that our single-window comparison cannot reveal. Second, proposal positions were predefined through a pilot study (see \autoref{sec:study}) rather than generated by an adaptive algorithm. This served to isolate effects of the visualization, but it leaves open how such techniques perform when proposals are produced dynamically, e.g., by the optimization-based adaptations of \citet{10.1145/3332165.3347945} or the Pareto-optimal sets of \citet{10.1145/3586183.3606799}. 

The preference for \cManualFull{} may therefore partly reflect the constraint of choosing from a fixed set, rather than a principled rejection of proposal-based interaction. Although 20 participants found the offered positions reachable and visible (\autoref{app:codebook}), 11 participants still called the proposal techniques restrictive, and the most-endorsed code ($n = 20$) was the wish to combine proposals with manual fine-tuning. It may also reflect a legacy bias, as the correlation between prior 2D window-management experience and \cManualFull{} preference (\autoref{app:familiarity}) is consistent with such an effect. A follow-up study contrasting fixed and algorithmically generated proposals could disentangle these explanations.

Third, our task introduces windows in a fixed sequence aligned with the trip-planning workflow, so the rearrangement trigger is window introduction rather than user-initiated context change; rearrangement under shifting priorities remains a complementary scenario that future work should explore.

Beyond these deliberate simplifications, several other limitations apply: Our seated, desk-based setup with a fixed environment leaves open how the comparison translates to standing or mobile scenarios, where the \cMiniatureFull may show stronger benefits for out-of-view proposals. In \cManualFull, some participants reported having to correct for inadvertent window rotations, which may have inflated the count of layout changes; restricting window orientation in future replications could avoid this confound. Instructing participants to prioritize arrangement over trip content was necessary to elicit layouting behavior, but likely inflated re-layouting frequency and may have shaped preference ratings. Further, in \cIconFull, we used the same icon for all windows (in line with prior work \cite{10.1145/3586183.3606799}), intentionally limiting the level of detail; content-specific icons are an obvious refinement for future work.

Evidently, our sample comprised primarily young, tech-savvy university students with moderate-to-high prior VR/AR exposure and high familiarity with 2D window management; the strong preference for \cManualFull may partly reflect this familiarity (legacy-bias) and may not generalize to the wider population. We used the raw (unweighted) NASA TLX, a validated and comparably sensitive alternative~\cite{hart2006nasa}; however, a weighted TLX might have better captured individual differences across the six dimensions. Finally, in our trial-order analysis (see \autoref{app:learningeffect}) we found no learning effects on arrangement-specific measures: overall task completion time showed an early-trial learning pattern that the Latin square design distributes evenly across conditions.

Several research directions emerge from this work: For multi-window arrangement, semantic window groups or saved layout presets may offer tractable approaches; a grid pattern around the user — where the environment acts as a placeholder space — presents another direction. User-defined layout presets, endorsed by 18 participants as a complement to proposal selection, represent another promising direction, supporting user-curated layouts for recurring tasks.
Beyond predefined positions, multi-objective optimization with a posteriori articulated preferences \cite{10.1145/3586183.3606799, 10.1145/3746059.3747645} would enable dynamic proposal generation tailored to the user's task and environment. \cMiniatureFull specifically warrants follow-up in scenarios with out-of-view window positions, where it
might offer greater benefits. 
Adding content-aware iconography to \cIconFull and supporting alternative interaction modalities (ray-based, gaze, snapping grids) would further broaden our design space. Finally, an AI-driven layer that detects the need for rearrangement and offers proposals unprompted would close the gap toward context-responsive systems — at which point the question of how to visualize proposals arguably becomes critical.

\section{Conclusion}
\label{sec:conclusion}

We compared three proposal-visualization techniques for window placement in MR (\cIconFull, \cWindowFull, and \cMiniatureFull) against a \cManualFull baseline in a within-subjects study where 24 participants placed seven windows for a trip-planning task in VR. Although \cIconFull and \cWindowFull reduced layouting time, participants consistently preferred direct manual adjustment, citing perceived control, familiarity, and the cognitive cost of selecting among predefined options. We outline implications for hybrid approaches that combine proposal-based suggestion with manual refinement.

\section{Open Science Statement}
Supplementary files include anonymized data, study instruments, and the codebook (Appendix~\ref{app:codebook}).

\begin{acks}
This research was funded by the \emph{Deutsche Forschungsgemeinschaft} (DFG, German Research Foundation) -- Project-ID~251654672 -- TRR~161.
\end{acks}

\bibliographystyle{ACM-Reference-Format}
\bibliography{sample-base}

\newpage
\clearpage

\appendix

\section {Full Pairwise Comparison Results}\label{app:pairwise}

The following tables report the complete pairwise comparisons
underlying the omnibus tests reported in \S{}5.1--\S{}5.3 of the
main paper. All $p$ values are Bonferroni-adjusted (raw $p \times k$,
capped at 1, where $k$ is the number of pairwise comparisons), and
all comparisons are tested against $\alpha = .05$. Effect sizes are
$r = |z|/\sqrt{N}$ with $N = 48$ for the non-parametric pairwise
contrasts ($n=24$ participants $\times$ 2 conditions). Condition
codes: MA = Manual Positioning, IP = Situated Icon Preview,
WP = Situated Window Preview, 3D = 3D Preview.

\subsection{Layouting Time}

\begin{table}[H]
\centering
\caption{Pairwise comparisons for layouting time (Dunn's test,
Bonferroni-adjusted).}
\label{tab:s2_1}
\begin{tabular}{lrrrrl}
\toprule
Comparison    & Test stat & $z$    & $p_{\text{adj}}$ & $r$ & Result \\
\midrule
WP vs.\ IP    &  0.333 &  0.894 & 1.000 & .13 & n.s. \\
WP vs.\ 3D    &  1.042 &  2.795 & \textbf{.031} & .40 & * \\
WP vs.\ MA    &  1.958 &  5.255 & \textbf{$<$.001} & .76 & ** \\
IP vs.\ 3D    & $-$0.708 & $-$1.901 &  .344 & .27 & n.s. \\
IP vs.\ MA    &  1.625 &  4.360 & \textbf{$<$.001} & .63 & ** \\
3D vs.\ MA    &  0.917 &  2.460 &  .083 & .36 & n.s. \\
\bottomrule
\end{tabular}
\end{table}

\subsection{Number of Layout Changes}

\begin{table}[H]
\centering
\caption{Pairwise comparisons for number of layout changes
(Dunn's test, Bonferroni-adjusted).}
\label{tab:s2_2}
\begin{tabular}{lrrrrl}
\toprule
Comparison    & Test stat & $z$    & $p_{\text{adj}}$ & $r$ & Result \\
\midrule
WP vs.\ IP    &  0.125 &  0.335 & 1.000 & .05 & n.s. \\
WP vs.\ 3D    &  0.125 &  0.335 & 1.000 & .05 & n.s. \\
WP vs.\ MA    &  2.000 &  5.367 & \textbf{$<$.001} & .77 & ** \\
IP vs.\ MA    &  1.875 &  5.031 & \textbf{$<$.001} & .73 & ** \\
3D vs.\ MA    &  1.875 &  5.031 & \textbf{$<$.001} & .73 & ** \\
IP vs.\ 3D    &  0.000 &  0.000 & 1.000 & .00 & n.s. \\
\bottomrule
\end{tabular}
\end{table}

\subsection{Overall Task Completion Time}

\begin{table}[H]
\centering
\caption{Pairwise comparisons for overall task completion time
(EMMEANS contrasts, Bonferroni-adjusted). A positive mean difference
(MD, in seconds) indicates the first (I) condition is slower than the
second (J).}
\label{tab:s2_3}
\resizebox{\columnwidth}{!}{%
\begin{tabular}{lrrrlc}
\toprule
Comparison    & MD (s)   & SE    & $p_{\text{adj}}$ & 95\% CI                  & Result \\
\midrule
MA vs.\ IP    & $+$10.10 & 29.47 & 1.000 & [$-$74.95, 95.15]    & n.s. \\
MA vs.\ 3D    & $-$56.39 & 26.34 &  .259 & [$-$132.41, 19.63]   & n.s. \\
MA vs.\ WP    & $+$33.87 & 26.74 & 1.000 & [$-$43.31, 111.05]   & n.s. \\
IP vs.\ 3D    & $-$66.49 & 29.30 &  .198 & [$-$151.06, 18.08]   & n.s. \\
IP vs.\ WP    & $+$23.78 & 25.09 & 1.000 & [$-$48.63, 96.18]    & n.s. \\
3D vs.\ WP    & $+$90.27 & 26.26 & \textbf{.013} & [14.47, 166.06]  & * \\
\bottomrule
\end{tabular}%
}
\end{table}

\subsection{UEQ Pairwise (Significant Subscales)}\label{app:b-ueq}
\noindent\textbf{Dependability} ($\chi^2(3) = 10.82$, $p = .013$, $W = .15$)

\begin{table}[H]
\centering
\begin{tabular}{lrrrrl}
\toprule
Comparison    & Test stat & $z$       & $p_{\text{adj}}$ & $r$ & Result \\
\midrule
3D vs.\ IP    &  0.625    &  1.677    &  .561  & .24 & n.s. \\
3D vs.\ WP    & $-$0.833  & $-$2.236  &  .152  & .32 & n.s. \\
3D vs.\ MA    &  1.125    &  3.019    & \textbf{.015}  & .44 & * \\
IP vs.\ WP    & $-$0.208  & $-$0.559  & 1.000  & .08 & n.s. \\
IP vs.\ MA    &  0.500    &  1.342    & 1.000  & .19 & n.s. \\
WP vs.\ MA    &  0.292    &  0.783    & 1.000  & .11 & n.s. \\
\bottomrule
\end{tabular}
\end{table}

\noindent\textbf{Efficiency} ($\chi^2(3) = 11.28$, $p = .010$, $W = .16$)

\begin{table}[H]
\centering
\begin{tabular}{lrrrrl}
\toprule
Comparison    & Test stat & $z$       & $p_{\text{adj}}$ & $r$ & Result \\
\midrule
3D vs.\ WP    & $-$0.208  & $-$0.559  & 1.000  & .08 & n.s. \\
3D vs.\ IP    &  0.813    &  2.180    &  .176  & .31 & n.s. \\
3D vs.\ MA    &  1.063    &  2.851    & \textbf{.026}  & .41 & * \\
WP vs.\ IP    &  0.604    &  1.621    &  .629  & .23 & n.s. \\
WP vs.\ MA    &  0.854    &  2.292    &  .131  & .33 & n.s. \\
IP vs.\ MA    &  0.250    &  0.671    & 1.000  & .10 & n.s. \\
\bottomrule
\end{tabular}
\end{table}

\noindent\textbf{Perspicuity} ($\chi^2(3) = 17.43$, $p < .001$, $W = .24$)

\begin{table}[H]
\centering
\begin{tabular}{lrrrrl}
\toprule
Comparison    & Test stat & $z$       & $p_{\text{adj}}$ & $r$ & Result \\
\midrule
3D vs.\ IP    &  0.917    &  2.460    &  .083  & .36 & n.s. \\
3D vs.\ WP    & $-$1.104  & $-$2.963  & \textbf{.018}  & .43 & * \\
3D vs.\ MA    &  1.396    &  3.745    & \textbf{.001}  & .54 & ** \\
IP vs.\ WP    & $-$0.188  & $-$0.503  & 1.000  & .07 & n.s. \\
IP vs.\ MA    &  0.479    &  1.286    & 1.000  & .19 & n.s. \\
WP vs.\ MA    &  0.292    &  0.783    & 1.000  & .11 & n.s. \\
\bottomrule
\end{tabular}
\end{table}

\subsection{Subjective Rating}

\begin{table}[H]
\centering
\caption{Pairwise comparisons for subjective preference rating
(Dunn's test, Bonferroni-adjusted).}
\label{tab:s2_5}
\begin{tabular}{lrrrrl}
\toprule
Comparison    & Test stat & $z$       & $p_{\text{adj}}$ & $r$ & Result \\
\midrule
3D vs.\ IP    &  0.250    &  0.671    & 1.000  & .10 & n.s. \\
3D vs.\ WP    & $-$0.292  & $-$0.783  & 1.000  & .11 & n.s. \\
3D vs.\ MA    &  1.458    &  3.913    & \textbf{$<$.001} & .57 & ** \\
IP vs.\ WP    & $-$0.042  & $-$0.112  & 1.000  & .02 & n.s. \\
IP vs.\ MA    &  1.208    &  3.242    & \textbf{.007}  & .47 & * \\
WP vs.\ MA    &  1.167    &  3.131    & \textbf{.010}  & .45 & * \\
\bottomrule
\end{tabular}
\end{table}

\noindent \emph{Significance markers used throughout this appendix: ** $p < .01$;
* $p < .05$ (Bonferroni-adjusted).}

\section {Learning Effects}
\label{app:learningeffect}
\begin{figure*}[h]
 \centering
     \includegraphics[width=.8\linewidth]{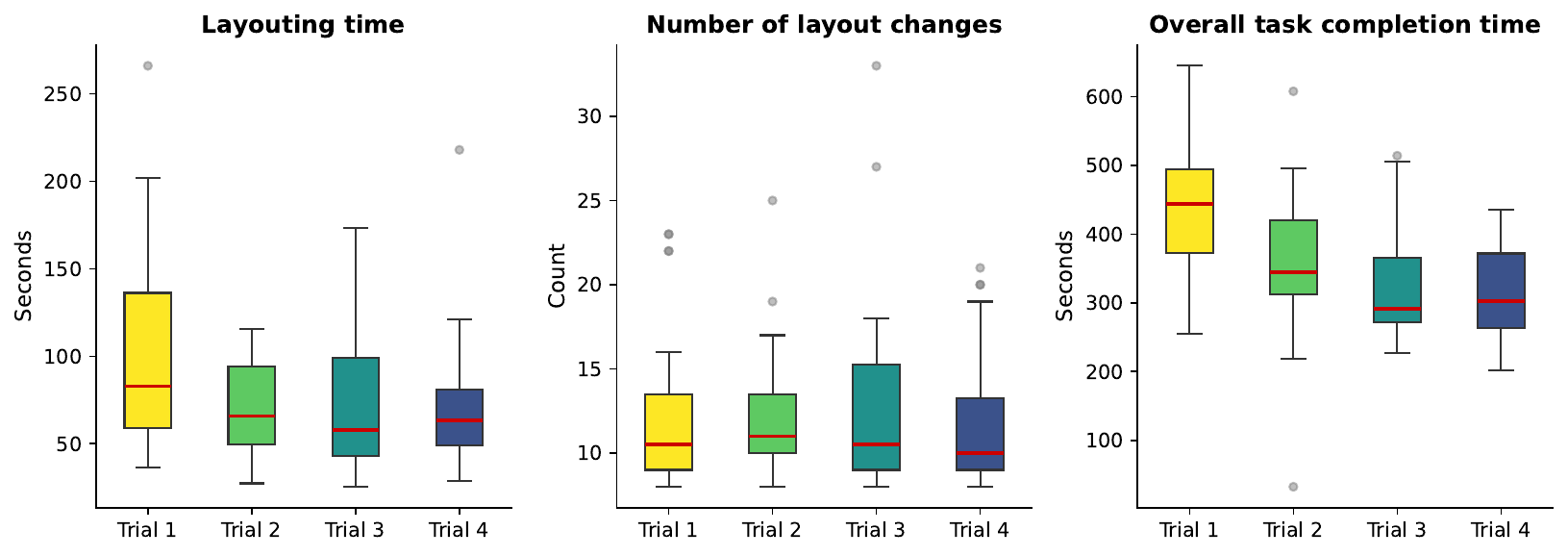}
     \caption{Distributions of three dependent measures across trial position (1st–4th session within each participant), aggregated over conditions.}
     \Description{Three panels, one per dependent measure, showing the distribution of each measure across trial position (first to fourth session within each participant), aggregated over all conditions, to check for learning and order effects.}
     \label{fig:learning}
 \end{figure*}

Three participants reported a perceived learning effect across conditions. To verify that strategy formation across trials does not confound the condition-level comparisons, we conducted complementary analyses treating trial position (1st, 2nd, 3rd, 4th session) as the independent variable; the Latin square design ensures condition and trial position are orthogonal. Friedman's ANOVA showed no significant effect of trial position on layouting time ($\chi^2(3) = 3.65$, $p = .302$, $W = .05$) or on the number of layout changes ($\chi^2(3) = 0.91$, $p = .822$, $W = .01$). For overall task completion time there was a significant effect ($\chi^2(3) = 25.75$, $p < .001$, $W = .36$): pairwise Wilcoxon signed-rank comparisons (Bonferroni-adjusted) showed Trial~1 significantly slower than Trials~2 ($z = -2.86$, $p = .026$, $r = .41$), 3 ($z = -3.74$, $p = .001$, $r = .54$), and 4 ($z = -3.57$, $p = .002$, $r = .52$); Trials~2--4 did not differ significantly as in \autoref{fig:learning}. Coefficients of variation showed no consistent trend across trial position. The pattern indicates that the learning visible in task completion time reflects familiarity with the trip-planning task itself rather than with the arrangement techniques: the two arrangement-specific measures showed no trial-position effect, while overall task completion plateaued after the first session. Because the Latin square design ($4! = 24$ participants) distributes any residual order effects evenly across conditions, the condition-level comparisons reported earlier are not confounded by this learning pattern.

\section {Study Power Analysis}
\label{app:poweranalysis}
Effect sizes are reported as partial $\eta^2$ for ANOVA, Kendall's $W$ for Friedman tests, and $r = |Z|/\sqrt{N}$ for pairwise non-parametric contrasts.
With $N = 24$ in a within-subjects design, our analyses were powered to detect medium-to-large effects. Post-hoc sensitivity analysis (computed in G*Power) indicates approximately $80\%$ power to detect Kendall's $W \geq .16$ in Friedman omnibus tests ($\alpha = .05$, $df = 3$) and pairwise effect sizes of $r \geq .50$ ($d_z \approx 0.73$) for Wilcoxon signed-rank comparisons after Bonferroni correction (six comparisons, $\alpha_{\mathrm{adj}} = .0083$). The main effects we report ($W = .45$ for layouting time, $W = .60$ for number of layout changes, $W = .83$ for stage-level layout changes, and $\eta_p^2 = .15$ for overall task completion time) substantially exceed these thresholds. Smaller observed effects --- including the UEQ Stimulation ($W = .11$) and Attractiveness ($W = .13$) omnibuses, which sit below the detection threshold --- should be interpreted tentatively; we therefore restrict claims to contrasts that survive correction at $\alpha = .05$.*

\section {Thematic Analysis Procedure}
\label{app:thhematicprocedure}

We analyzed the interview transcripts following Braun and Clarke's reflexive thematic analysis \cite{braun2006using}. Interviews were audio-recorded and transcribed. Two authors carried out the analysis: a primary coder and a second coder for validation. In the familiarization phase, the primary coder read all transcripts and noted initial impressions. During initial coding, meaningful segments were identified and assigned descriptive codes using MAXQDA and spreadsheets; coding was iterative, with codes added, merged, and refined as new transcripts were processed. Example initial codes included ``put it wherever I want'', ``fine adjustments'', ``multi-step selection'', ``icon image looks unfamiliar'', ``visual clutter'', and ``creative''.

Related codes were clustered into candidate themes — for example, codes about freedom of placement, fine adjustment, and resistance to predefined positions coalesced into a Perceived Control theme; codes about simplicity and quickness clustered into a Cognitive Cost of Selection theme. The validating second coder reviewed candidate codes and themes, and the two coders refined the structure through discussion: some candidates were merged (separate codes about visual clutter in \cWindowFull and small representations in \cIconFull folded into a single Proposal Informativeness theme), others were split, and a small number of codes were dropped where they did not pattern across participants. Final themes were named to capture the patterned meaning each represented and to address our research questions. We did not compute formal inter-rater reliability, in keeping with the reflexive thematic analysis position that consensus and researcher reflexivity, rather than agreement metrics, are appropriate quality indicators for interpretive analysis. Four themes emerged: Perceived Control over Placement, Cognitive Cost of Selecting from Proposals, Familiarity and Visual Recognition, and Informativeness of the Proposal.

\section{Familiarity and Preference}
\label{app:familiarity}
Table \ref{tab:familiarity} relates participants' prior experience to their preference ratings. Experience with \textbf{2D window management} correlates with a stronger preference for \cManualFull{} over the proposal techniques ($\rho = +.40$, $p = .051$; marginal) and a weaker preference for \cMiniatureFull{} ($\rho = -.42$, $p = .043$). Participants with more VR/AR and 3D UI experience rated \cIconFull{} lower ($\rho = -.46$, $p = .022$; $\rho = -.52$, $p = .009$). These correlations are exploratory and based on 24 participants; they are indicative rather than confirmatory.
\begin{table}[H]
\centering
\resizebox{\columnwidth}{!}{%
\begin{tabular}{lrrrrr}
\toprule
Prior experience & MA & IP & 3D & WP & MA vs.\ proposals \\
\midrule
2D window management  & $+.28$ & $-.27$ & $-.42^{*}$  & $+.26$ & $+.40$ \\
VR/AR head-mounted    & $-.16$ & $-.46^{*}$  & $-.19$ & $-.11$ & $+.20$ \\
3D user interfaces    & $-.16$ & $-.52^{**}$ & $-.07$ & $+.02$ & $+.09$ \\
\bottomrule
\end{tabular}%
}
\caption{Spearman correlations between prior experience and preference ($N = 24$). MA~=~\cManualFull{}, IP~=~\cIconFull{}, 3D~=~\cMiniatureFull{}, WP~=~\cWindowFull{}. The last column compares MA against the average of the three proposal techniques. $^{*}p < .05$, $^{**}p < .01$.}
\label{tab:familiarity}
\end{table}

\section{Detailed Task Flow}
\label{app:taskflow}

\begin{figure}[H]
  \centering
  \includegraphics[width=0.91\textwidth]{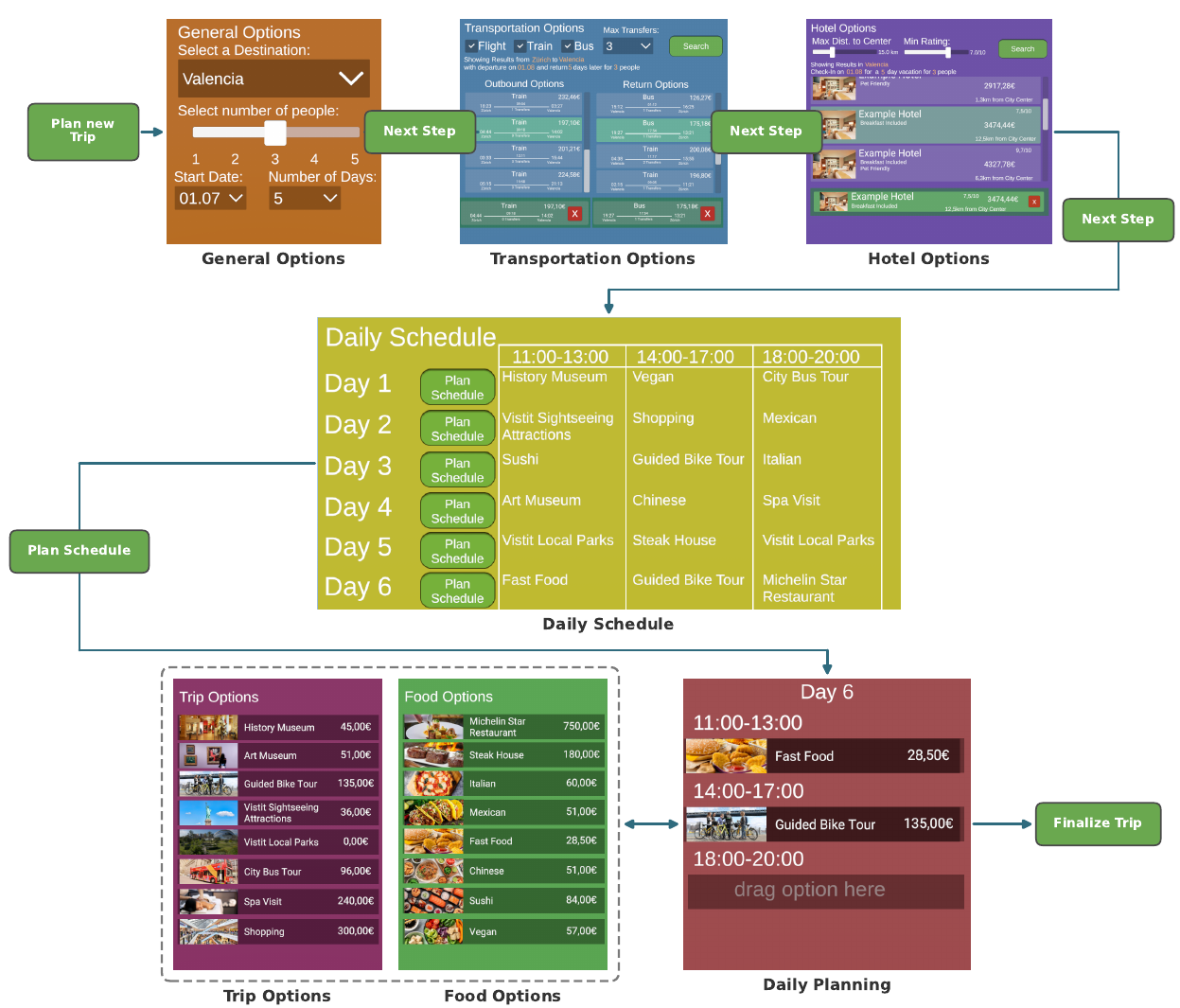}
  \caption{Detailed view of the trip-planning task. Each stage introduces one new window, which participants place before continuing. Green buttons advance the task; the \textit{Finalize Trip} button ends it.}
  \Description{A detailed walkthrough of the trip-planning interface arranged in three rows of screenshots. The top row shows the setup windows (General Options, Transportation Options, and Hotel Options) linked by ``Next Step'' buttons. The middle row shows the Daily Schedule window. The bottom row shows the Daily Planning window linked to a framed pair of Trip Options and Food Options windows, ending with a ``Finalize Trip'' button. Green buttons advance the task between stages.}
  \label{fig:detailedflow}
\end{figure}

Figure \ref{fig:detailedflow} shows the complete interface of the trip-planning task: the content of each of the seven windows, the green buttons that trigger the next stage, and the transitions between stages.

\onecolumn

\balance
\onecolumn

\section{Codebook}
\label{app:codebook}

The table below lists the codes from our reflexive thematic analysis, grouped by sub-theme and theme. \textit{n} indicates the number of
participants (out of 24) whose interview was coded with the respective code. The final group, \textit{Additional codes outside the four themes}, collects codes that recurred in the data but did not coalesce into one of the four visualization-design themes reported in \autoref{sec:interview}; the most salient of these (\textit{Suitability of Positions}) is discussed separately in \autoref{sec:discussion}.

\bigskip
\noindent
\setlength{\tabcolsep}{6pt}
\renewcommand{\arraystretch}{1.15}
\begin{tabular}{p{0.20\linewidth} p{0.62\linewidth} r}
\toprule
\textit{Sub-theme} & \textit{Code} & \textit{n} \\
\midrule
\multicolumn{3}{l}{\textbf{Theme 1: Perceived Control over Placement}} \\
\addlinespace[2pt]
User Control
  & Combine proposals with manual fine-tuning & 20 \\
  & Manual: ``Put it wherever I want'' & 19 \\
  & Manual: Fine-grained adjustments & 15 \\
  & Manual: Freedom of placement & 14 \\
  & Proposals felt restrictive & 11 \\
Orientation
  & Manual: Free rotation was annoying & 10 \\
\addlinespace[4pt]
\multicolumn{3}{l}{\textbf{Theme 2: Cognitive Cost of Selecting from Proposals}} \\
\addlinespace[2pt]
Ease of Use
  & Icon Preview: Simple and intuitive & 13 \\
  & Window Preview: Convenient and simple & 7 \\
  & 3D Preview: Hard to understand & 7 \\
  & 3D Preview: Hard to navigate between proposals & 5 \\
  & 3D Preview: Direction arrow misleading & 4 \\
  & Icon Preview: Easy and quick & 3 \\
  & Proposals: Quicker than manual & 3 \\
Effort
  & 3D Preview: Multi-step selection & 8 \\
  & Manual: Too much effort & 5 \\
\addlinespace[4pt]
\multicolumn{3}{l}{\textbf{Theme 3: Familiarity and Visual Recognition}} \\
\addlinespace[2pt]
Familiarity
  & 3D Preview: Creative / novel & 6 \\
  & Icon Preview: Icon image looked unfamiliar & 5 \\
\addlinespace[4pt]
\multicolumn{3}{l}{\textbf{Theme 4: Informativeness of the Proposal}} \\
\addlinespace[2pt]
Level of Detail
  & Window Preview: Shows exact appearance & 9 \\
  & 3D Preview: Conveys relative layout & 5 \\
  & Icon Preview: Conveys position only & 4 \\
Proposal Size
  & Window Preview: Visual clutter at scale & 6 \\
  & Icon Preview: Less visual clutter & 4 \\
  & Icon Preview: Representation too small & 2 \\
  & 3D Preview: Spheres too small & 2 \\
\addlinespace[4pt]
\multicolumn{3}{l}{\textbf{Additional codes outside the four themes}} \\
\addlinespace[2pt]
Suitability of Positions
  & Positions: Reachable and visible & 20 \\
  & Position: Dominant-hand placement preferred & 4 \\
  & Visualization: Grid pattern around the user & 2 \\
  & Position: Group windows semantically & 2 \\
  & Position: Single center position not feasible & 1 \\
Personalization
  & Configure reusable layouts for repetitive tasks & 18 \\
Learning
  & The trip planning procedure became easier over time & 5 \\
\bottomrule
\end{tabular}

\end{document}